\documentclass[a4paper,11pt]{article}
\usepackage{pos}
\usepackage{cleveref}

\newcommand{\BE}{\begin{equation}}
\newcommand{\EE}{\end{equation}}

\newcommand{\sqrtsgaga}{\sqrt{s_{\gamma\gamma}}}
\newcommand{\sgaga}{s_{\gamma\gamma}}

\title{Probing the Higgs potential at a Photon Collider}

\author[a]{Marten Berger}
\author[b]{Johannes Braathen}
\author*[a,b]{Gudrid Moortgat-Pick}
\author[a,b]{Georg Weiglein}

\affiliation[a]{II. Institut für Theoretische Physik, Universität Hamburg, Luruper Chaussee 149, 22761 Hamburg, Germany}

\affiliation[b]{ Deutsches Elektronen-Synchrotron DESY, Notkestr.~85, 22607 Hamburg, Germany }

\emailAdd{johannes.braathen@desy.de}
\emailAdd{marten.berger@desy.de}
\emailAdd{gudrid.moortgat-pick@desy.de}
\emailAdd{georg.weiglein@desy.de}

\abstract{   
A $\gamma\gamma$ collider, either in conjunction with an $e^+e^-$ linear collider or as a stand-alone facility, offers a very attractive Higgs physics programme at relatively low centre-of-mass (c.m.) energies. While the Higgs boson that has been discovered at the LHC can be studied in detail in resonant production at 125~GeV, a c.m.\ energy as low as 280~GeV can probe the Higgs potential via the Higgs pair production process providing access to the trilinear Higgs-boson self-coupling.  
High polarisation of the photon beams (produced via Compton back-scattering) can be achieved and adjusted by flipping the polarisation of the incident laser. 
The prospects for exploring the Higgs pair production process at a $\gamma\gamma$ collider are assessed by
comparing different running scenarios utilising different types of the incident laser. 
The possibility to use photon polarisations for disentangling different 
kinds of contributions to the Higgs pair production process
is emphasised. 
}

\FullConference{The European Physical Society Conference on High Energy Physics (EPS-HEP2025)\\
7-11 July 2025\\
Marseille, France\\}

\begin{document}

\begin{flushright}
\texttt{DESY-25-133}
\end{flushright}

\maketitle


\section{Photon--photon collider: basic features}

A photon--photon ($\gamma\gamma$) collider 
operates by converting high-energy electrons into high-energy photons via Compton back-scattering: laser photons with 
energy $\omega_0$ collide with the electrons of energy $E_0$ at a conversion point, a short distance $b$ before the interaction point, see for example fig.\ 1 in Ref.~\cite{Telnov:2020gwc}. Real (i.e.\ on-shell)
high-energy photons are scattered in the direction of the interaction point. 
This setup can be used at 
an $e^+e^-$-collider to enable the additional modes of $\gamma\gamma$ and $\gamma e$ collisions, with luminosities and energies comparable to those of $e^+e^-$ collisions. 
Furthermore, a $\gamma\gamma$ collider can also be built based on an $e^-e^-$ collider.
The main parameter for photon colliders 
in both cases
is the dimensionless quantity $x$ describing the laser--$e$ collision, which is given by~\cite{Ginzburg}\footnote{We note that we are using here natural units (i.e.\ $c=1$). If we had not done so, $x$ would contain an additional factor $1/c^{4}$, as e.g.\ in Refs.~\cite{Ginzburg}.}
\begin{equation}
  x = \frac{4E_0\omega_0}{m_e^2}\cos^2\frac{\theta}{2}\simeq  15.3 \left[\frac{E_0}{\text{TeV}}\right]\left[\frac{\omega_0}{\text{eV}}\right]=19\left[\frac{E_0}{\text{TeV}}\right]\left[\frac{\mu\text{m}}{\lambda}\right],
    \label{eq:photon_collider}
\end{equation}
where $\theta$ is the angle between the laser and electron beam. The maximum energy $\omega_{\mathrm{max}}$ that a scattered photon can reach is then given by $\omega_{\mathrm{max}} = \frac{x}{x+1}E_0$. Historically, an upper bound of $x<4.8$ was chosen as for 
higher values of $x$ the Breit-Wheeler process, and for $x>8.0$ also the Bethe-Heitler process, 
would drastically decrease the luminosity.
With 
this restriction a photon collider would be able to achieve 
centre-of-mass (c.m.) energies up to $80\%$ of the c.m.\ energy of the corresponding $ee$-collider.

\section{Possibilities for the implementation of a $\gamma\gamma$-interaction region}
Recently a new design with $x\geq1000$ has been discussed, 
yielding photon collider
energies close to $100\%$ of the $ee$-collider energy, by using a XFEL-laser for the Compton back-scattering process~\cite{Barklow:2022vkl,Barklow:2023ess}. It has been shown that going so far beyond the cut-off at $4.8$ $(8.0)$ the resulting luminosities are still significant and even feature a narrower peak around the maximum energy~\cite{Barklow:2024XCC}. 
Therefore, two different types of $\gamma\gamma$-collider setups can now be considered: 
\begin{itemize}
    \item the optical-laser setup, with $x<4.8$; and
    \item the XFEL-based setup, or XCC, with $x\geq1000$.
\end{itemize}
Both are capable of offering a rich physics programme with high-energy photons, either as a stand-alone $\gamma\gamma$-collider or an addition to any $ee$-collider, as will be discussed in the following for the example of 
the Higgs pair production process.

\section{Higgs pair production at a $\gamma\gamma$ collider}
We investigate here the possibility of pair-producing Higgs bosons at different options of $\gamma\gamma$ colliders~\cite{ourpaper,slacpaper}. 
In this context, we assess the sensitivity 
for probing
the trilinear self-coupling of the detected Higgs boson or, equivalently, its coupling modifier $\kappa_\lambda$. The collider-level cross-section for $\gamma\gamma\to hh$, 
where $h$ denotes the detected Higgs boson at 125~GeV,
is given by 
\begin{equation}
    \sigma = \int_{4m_h^2/s}^{y^2_{max}}\text{d}\tau\frac{1}{2}\left[\frac{1}{L_{\gamma\gamma}^{++}}\frac{\text{d}L^{++}_{\gamma\gamma}}{\text{d}\tau}\hat{\sigma}_{++}(\sgaga)+\frac{1}{L_{\gamma\gamma}^{+-}}\frac{\text{d}L^{+-}_{\gamma\gamma}}{\text{d}\tau}\hat{\sigma}_{+-}(\sgaga)\right] .
    \label{eq:GammaGamma}
\end{equation}
Here $\hat\sigma_{\lambda_1\lambda_2}$ denote the cross-sections  for the photon polarisation configurations $\{\lambda_1,\lambda_2\}$ (
with $\lambda_i=\pm$) at the partonic level (i.e.\ in the photon-photon system), $L_{\gamma\gamma}^{\lambda_1\lambda_2}$ are the corresponding luminosity spectra, and $\sqrtsgaga$ is the 
c.m.\ energy of the colliding photons. The integration variable is defined as the fraction $\tau\equiv\sgaga/ s$, while the upper integration limit $y_{max}$ is the maximum energy fraction $y_{max}\equiv\omega_{max}/E_0$. 
In the lower integration bound and the definition of $\tau$, $s$ denotes the squared 
c.m.\ energy of the $e^+e^-$ or $e^-e^-$
collider. 

\subsection{Partonic-level cross-sections}
\begin{figure}
    \centering
    \includegraphics[height=1.5cm]{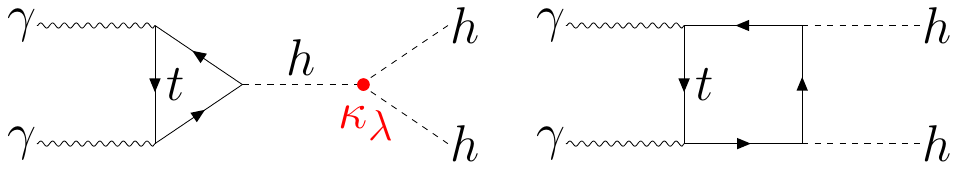}\\
    \includegraphics[height=1.5cm]{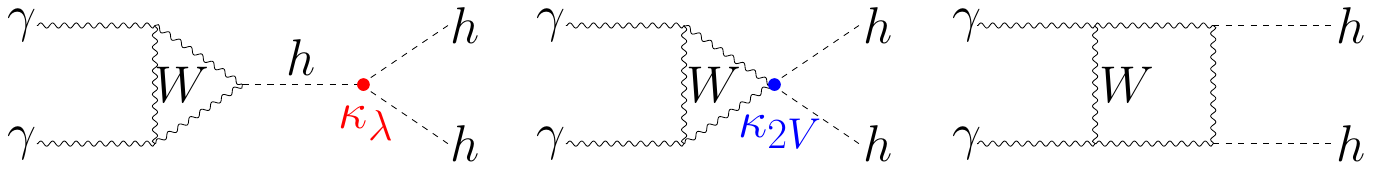}
    \caption{Representative Feynman diagrams for the process $\gamma\gamma\to hh$. The upper row shows diagrams involving top-quark loops, while the diagrams in the lower row correspond to gauge-sector contributions. }
    \label{fig:diagrams}
\end{figure}

Example diagrams contributing to the 
Higgs pair
production process are displayed in \cref{fig:diagrams}. Unlike the $gg\to hh$ process at the (HL-)LHC, where only coloured particles contribute 
in the loop
at leading order, there are additional contributions from the gauge sector at the same order for $\gamma\gamma\to hh$  (see the lower row of \cref{fig:diagrams}). 

\begin{figure*}
    \centering
    \includegraphics[width=0.47\textwidth]{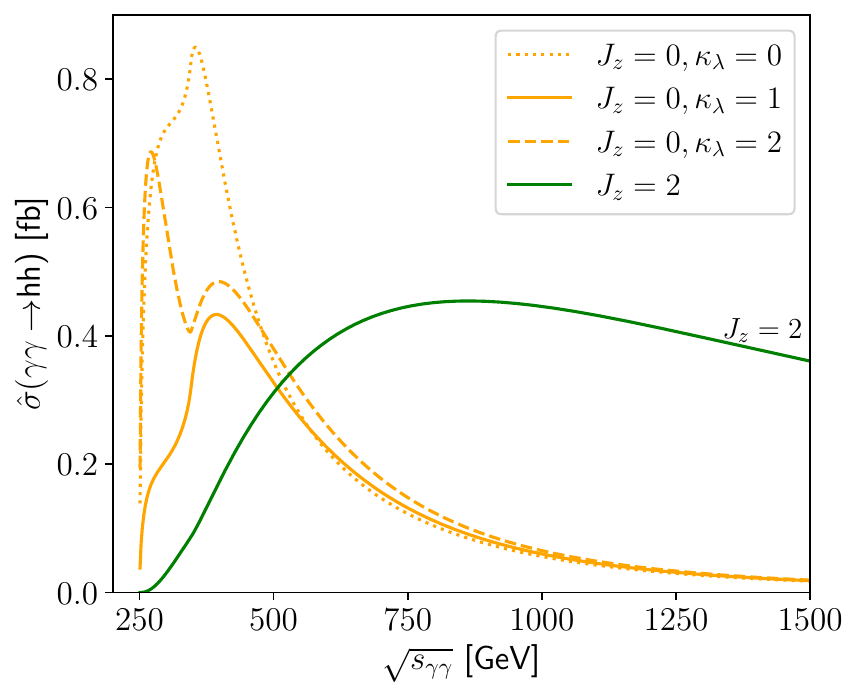}
    \includegraphics[width=0.52\textwidth]{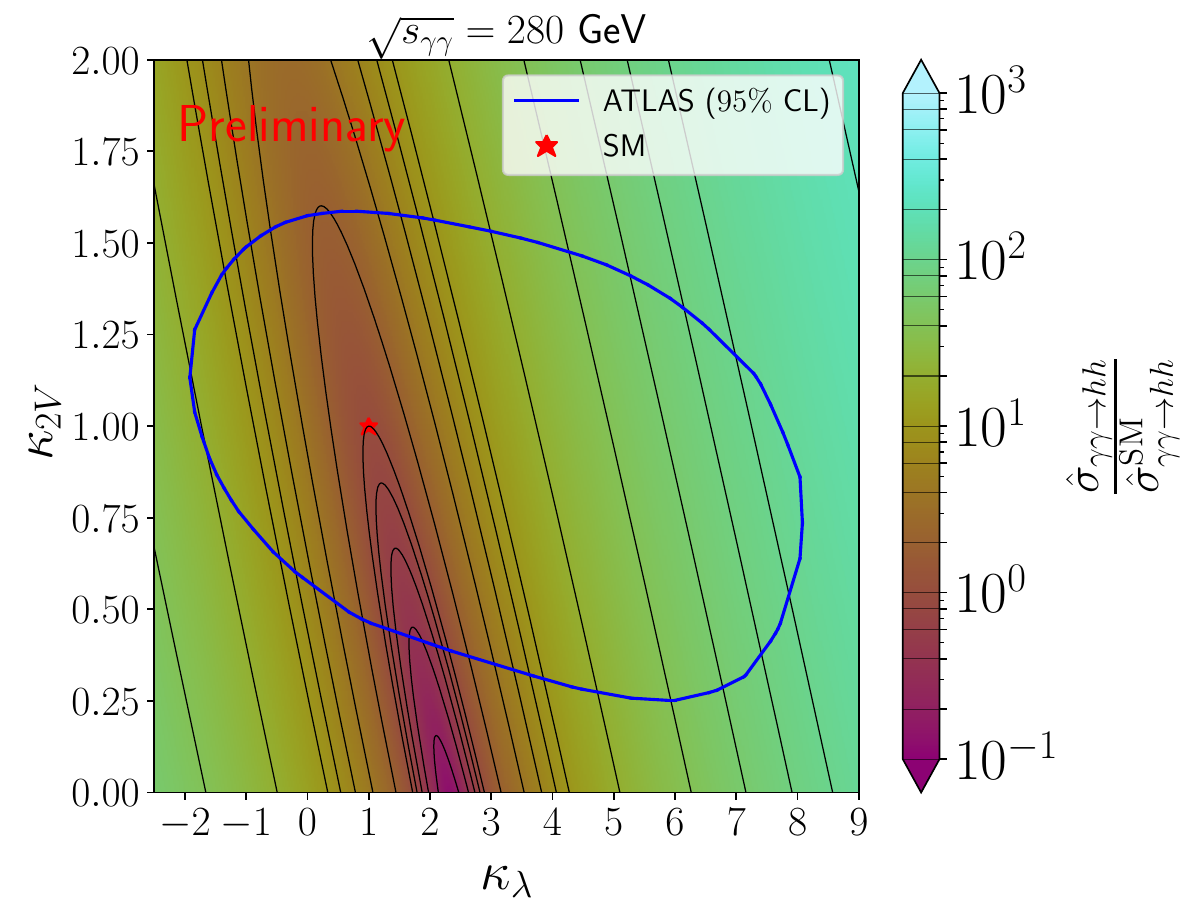}
    \caption{ 
    Higgs pair production cross section (at the partonic level) as a function of the photon-photon 
    c.m.\ energy $\sqrt{s_{\gamma\gamma}}$. The orange lines indicate results for $\hat\sigma_{++}$ (i.e.\ $J_z=0$) for different values of $\kappa_\lambda$, while the green line shows the result for $\hat\sigma_{+-}$ (i.e.\ $J_z=2$). \textit{Right:} Predictions for the unpolarised partonic cross-section for $\gamma\gamma\to hh$, defined as $(\hat{\sigma}_{++}+\hat{\sigma}_{+-})/2$, normalised to its value in the SM, shown as contours in the plane of $\kappa_\lambda$ and $\kappa_{2V}$, for $\sqrt{s_{\gamma\gamma}}=280\text{ GeV}$. The blue line indicates the current ATLAS limits at the 95\% C.L.~\cite{ATLAS:2024ish}.}
    \label{fig:partonic}
\end{figure*}

For our analysis 
we have rederived the leading-order (one-loop) results in the SM using \texttt{FeynArts}~\cite{Kublbeck:1990xc,Hahn:2000kx} and \texttt{FormCalc}~\cite{Hahn:1998yk,Hahn:2016ebn}, and found agreement both with Refs.~\cite{Jikia:1992mt,Asakawa:2008se,Asakawa:2010xj,Bharucha:2020bhy} and with the amplitudes used in \texttt{Whizard}~\cite{Kilian:2007gr,Moretti:2001zz}. 
Moreover, we obtained analytic expressions for $\hat\sigma(\gamma\gamma\to hh)$ 
for arbitrary values\footnote{Large deviations in $\kappa_\lambda$ from the SM value can occur in models with extended scalar sectors, due to radiative corrections from the BSM scalars, see e.g.\ Refs.~\cite{Kanemura:2004mg,Kanemura:2017wtm,Bahl:2022jnx,Biekotter:2022kgf,Bahl:2023eau,Braathen:2025qxf}.} of
$\kappa_\lambda$ and $\kappa_{2V}$ --- the latter being the coupling modifier of the interactions between two Higgs bosons and two gauge bosons or between two Higgs and two Goldstone bosons. In the left plot of \cref{fig:partonic}, we present our results for $\hat\sigma_{++}$ (orange) and $\hat\sigma_{+-}$ (green) as a function of 
$\sqrtsgaga$ --- noting that only the cross-section $\hat\sigma_{++}$ (i.e.\ for $J_z=0$) exhibits a dependence on $\kappa_\lambda$. We find 
that in the SM-like case (orange solid line), the cross-section for $J_z=0$ peaks around $\sqrtsgaga\simeq 400\text{ GeV}$. On the other hand, if one allows $\kappa_\lambda$ to vary, the largest 
deviations from the cross-section for the SM value of $\kappa_\lambda = 1$ 
occur for $\sqrtsgaga\simeq 280\text{ GeV}$. 
This is further illustrated
in the right plot of \cref{fig:partonic}, where we show the predictions for the unpolarised $\hat\sigma(\gamma\gamma\to hh)$ cross-section at $\sqrtsgaga\simeq 280\text{ GeV}$, normalised to its SM prediction, as contours in the plane of $\kappa_\lambda$ and $\kappa_{2V}$. Within the region allowed by the current ATLAS results~\cite{ATLAS:2024ish}, variations --- and in particular enhancements --- of several orders of magnitude are possible. Moreover, for the case $\kappa_{2V}=1$, the minimum of the cross-section variation with $\kappa_\lambda$ is located close to $\kappa_\lambda\simeq 1$; this contrasts with the cases of $e^+e^-\to Zhh$ at LCF550~\cite{LinearColliderVision:2025hlt,Berggren:2025fpw} and $gg\to hh$ at the (HL-)LHC~\cite{LHCHiggsCrossSectionWorkingGroup:2016ypw}, which exhibit minima at $\kappa_\lambda\simeq 1.5$ and $\kappa_\lambda\simeq 2$, respectively. The 
contour lines
of equal cross-section values signal a low degree of correlation between variations from $\kappa_\lambda$ and $\kappa_{2V}$, implying that with the expected constraints on $\kappa_{2V}$ after HL-LHC, to about $10\%$ accuracy, the remaining uncertainty on $\kappa_{2V}$ would not significantly degrade 
the accuracy of the determination of 
$\kappa_\lambda$ 
obtained from the results for the Higgs pair production process 
at a $\gamma\gamma$ collider with 280~GeV. 

\subsection{Luminosity Spectra}

\begin{figure*}[tbp]
  \centering
  \begin{minipage}[b]{0.325\linewidth}
    \includegraphics[width=\linewidth]{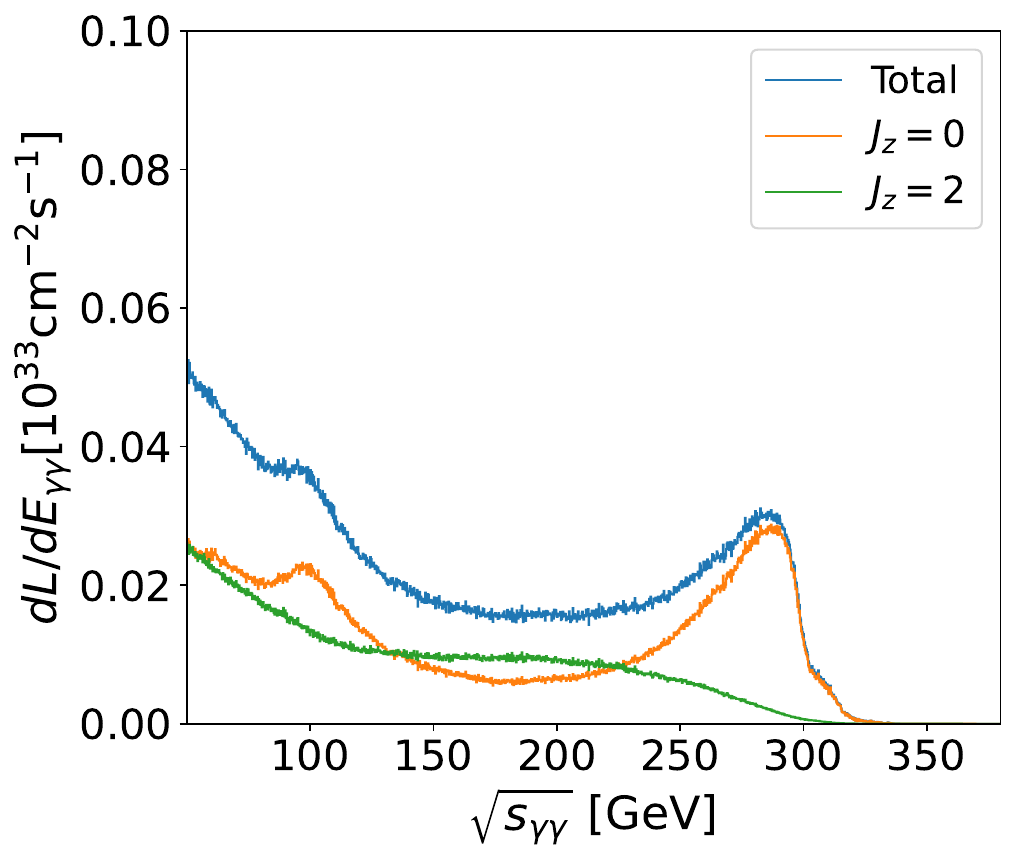}
    \label{fig:optical380}
  \end{minipage}
  \begin{minipage}[b]{0.317\linewidth}
    \includegraphics[width=\linewidth]{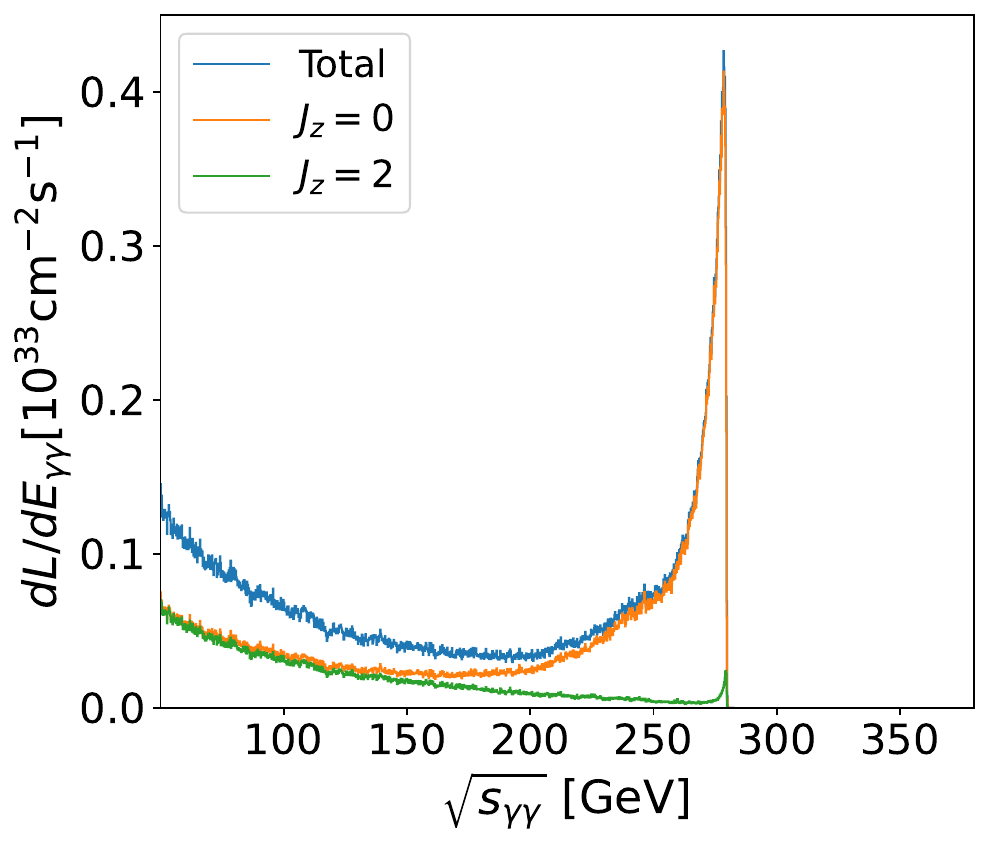}
    \label{fig:xcc280}
  \end{minipage}
  \begin{minipage}[b]{0.325\linewidth}
    \includegraphics[width=\linewidth]{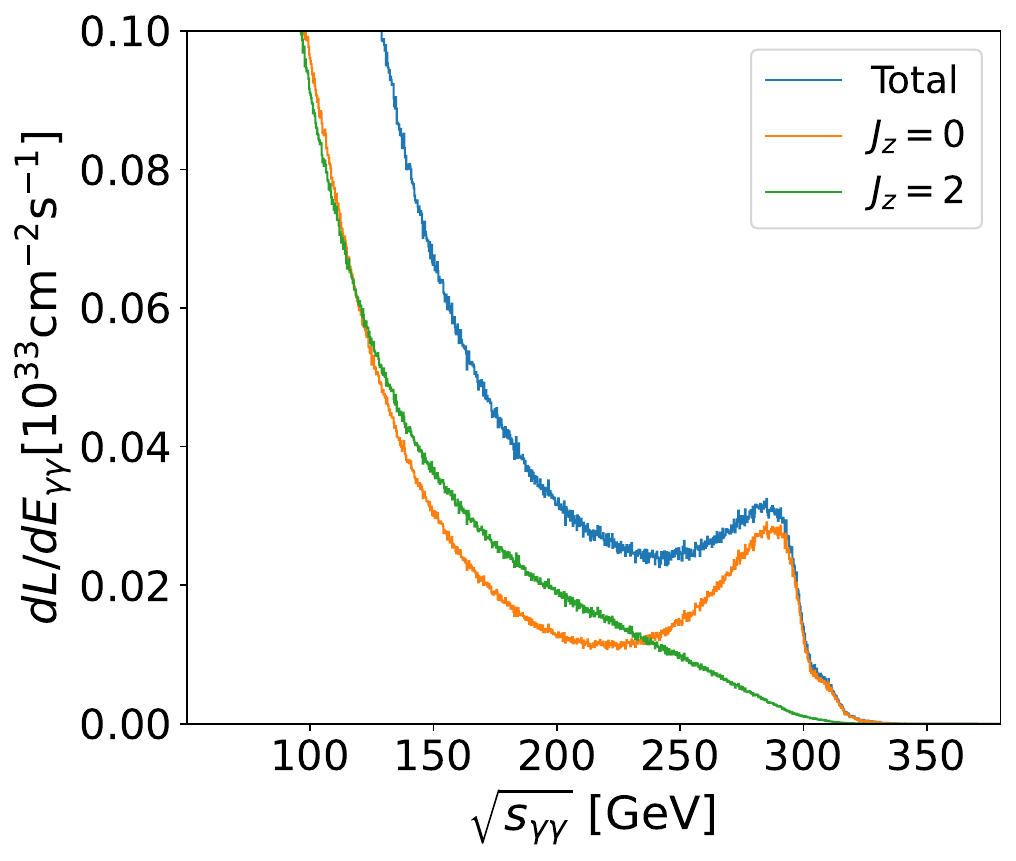}
    \label{fig:positron}
  \end{minipage}
  \vspace{-1.5em}
  \caption{
  The luminosity spectrum for the photon collider using an optical laser at a 380 GeV $e^-e^-$-collider (left), for the XCC at a 280 GeV $e^-e^-$-collider (centre) and for an optical photon collider at a 380 GeV $e^+e^-$-collider (right), showing the total (blue), $J_z=0$ (orange) and $J_z=2$ (green) luminosity spectra. Calculated with \texttt{CAIN} using a beam setup adapted from the ILC design~\cite{Bechtel:2006mr}, with the new parameters given in tables 23 and 24 of Ref.~\cite{LinearColliderVision:2025hlt}.}
  \label{fig:spectra}
\end{figure*}

In order to achieve the highest
sensitivity to the trilinear Higgs-boson self-coupling, the $\gamma\gamma$-collider setup should be optimized for the $J_z=0$ state. For this purpose, both lasers 
need to have the same polarisation, and the same is true for both $e$-beams. 
Moreover, the electron helicity $\lambda_e$ and photon circular polarisation $P_c$ need to be 
such that $\lambda_eP_c<0$ for the optical setup and $\lambda_eP_c>0$ for the XFEL-like setup (the latter in order to suppress $e^+e^-$ pair-production). While analytical expressions exist for the calculation of the luminosity spectrum for a collider with a given 
value of
$x$, these do not take all the beam and beam-beam interactions 
into account. Therefore, we have used the Monte-Carlo code \texttt{CAIN}~\cite{Chen:1994jt}, which includes Breit-Wheeler, Bethe-Heitler and non-linear QED processes, to obtain realistic luminosity spectra. The left and middle plots of \cref{fig:spectra} display the spectra computed with \texttt{CAIN} 
for an optical $\gamma\gamma$-collider based on a $380$ GeV $ee$-collider (left) and for the XFEL-based $\gamma\gamma$-collider 
at a $280$ GeV $ee$-collider (centre); both spectra have their maximum 
around $280$ GeV. 
In the past the photon collider option has mainly been discussed for $e^-e^-$-colliders, due to the low polarisation of $e^+$, but the progress in positron polarisation now opens the possibility 
to run the $\gamma\gamma$-collider in the $e^+e^-$ mode.
The corresponding 
luminosity spectrum is shown
in the right plot of \cref{fig:spectra}. It can be seen that 
compared to the other two cases many 
more low energy photons are produced, however around the maximum energy the spectrum is very close to the classical $e^-e^-$ mode setup. It thus appears 
possible to use this setup for any process above $200$~GeV.
Finally, we note that having $\gamma\gamma$ collisions at the second interaction region of an $e^+e^-$ LCF --- in parallel to $e^+e^-$ collisions at the first interaction region --- offers additional luminosity in $\gamma\gamma$ events, at possibly only a moderate cost (depending on the precise setup) in terms of the total $e^+e^-$ luminosity.  

\subsection{Integrated cross-section for Higgs pair production}

\begin{figure*}[tbp]
  \centering
 
    \includegraphics[width=0.5\linewidth]{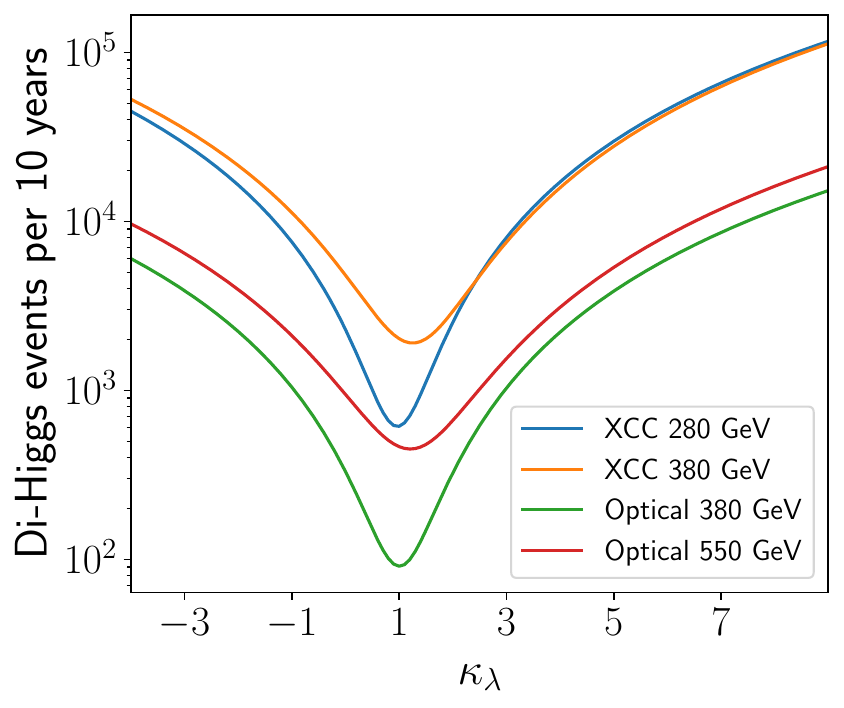}
    \caption{Number of 
    Higgs pair production events at different options of $\gamma\gamma$ colliders as a fuction of $\kappa_\lambda$. The blue and orange lines indicate the results for XCC options with $E_{e^-e^-}=280$~GeV 
    and 380~GeV, respectively. The green and red lines correspond to optical laser based colliders with $E_{e^-e^-}=$ 380~GeV and 550~GeV, respectively.}
    \label{fig:dihiggsnumevents}
\end{figure*}

We now combine our analytical results for $\hat\sigma(\gamma\gamma\to hh)$~\cite{ourpaper} with the luminosity spectra obtained with \texttt{CAIN} in order to obtain collider-level results for Higgs pair production. Taking into account the total integrated luminosities, we present in \cref{fig:dihiggsnumevents} the total number of 
Higgs pair production
events that could be produced per decade of run time as a function of $\kappa_\lambda$ (for $\kappa_{2V} = 1$) for different options of photon colliders. The blue and orange curves correspond to XFEL-based XCC options, at $E_{e^-e^-}=280\text{ GeV}$ (blue) and 380~GeV (orange), while the green and red curves display the results for optical-laser based colliders, with $E_{e^-e^-}=380\text{ GeV}$ (green) and 550~GeV (red). Overall, the XCC options 
provide a higher event rate for the Higgs pair production process. It amounts to
about one order of magnitude more events 
for the XFEL-based options as compared to
the optical-laser based ones of similar maximal energies of the colliding photons ($\sqrtsgaga$). 
While the Higgs pair production cross-section for the XCC at 280~GeV and the optical-laser based option utilising $E_{e^-e^-}=380\text{ GeV}$ has a minimum near the SM value of $\kappa_\lambda = 1$, as already noted for 
the partonic-level results, 
these photon collider options 
exhibit the strongest dependence on 
the trilinear Higgs-boson self-coupling for $\kappa_\lambda \neq 1$.
They will therefore provide a very precise determination of $\kappa_\lambda$ even for the case where the SM value is realised in nature, as a consequence of the 
stringent constraints on non-standard values of $\kappa_\lambda$ 
that would give rise to much enhanced cross-sections. 
We note that in Ref.~\cite{slacpaper} it has been demonstrated that the XCC at 280 GeV would enable 
the determination of
$\kappa_\lambda$ with a precision of about $5\%$ (comparable to FCC-$hh$) for most of the allowed range of $\kappa_\lambda$ except between about 0.5 and 1.5 (see also Ref.~\cite{LinearColliderVision:2025hlt}, as well as Ref.~\cite{Kawada:2012uy} for earlier work). 
A photon collider with rather moderate c.m.\ energy therefore has excellent prospects to very significantly improve the determination of the trilinear Higgs-boson self-coupling compared to the ultimate precision at the HL-LHC.

\section{Conclusions}

We have discussed the capabilities of a $\gamma\gamma$-collider to probe the Higgs potential via the Higgs pair production process, 
investigating the collider options utilising an XFEL (XCC) or an optical laser system for different c.m.\ energies~\cite{ourpaper}.
At the partonic level, we have 
analysed 
the dependence of $\sigma(\gamma\gamma\to hh)$ on the combined effects of the coupling modifiers $\kappa_\lambda$ and $\kappa_{2V}$.  
For $\sqrtsgaga=280\text{ GeV}$ the minimum of the cross-section for $\kappa_{2V}=1$ is located close to the SM value of $\kappa_\lambda=1$. 
The steep dependence of the cross section at $\sqrtsgaga=280\text{ GeV}$ on 
$\kappa_\lambda$ for values differing from the SM prediction implies 
that this $\gamma\gamma$ collider energy is particularly promising for a precise determination of the trilinear Higgs-boson self-coupling. 
We have furthermore demonstrated that the 
expected uncertainty on $\kappa_{2V}$ 
after the HL-LHC of about 10\% 
will 
not significantly degrade the 
accuracy of the determination of
$\kappa_\lambda$ from 
the Higgs pair production process at the photon collider.

While most studies for $\gamma\gamma$ colliders up to now have restricted  themselves to setups based on $e^-e^-$-colliders, we have also considered the possibility 
to operate a $\gamma\gamma$ collider in conjunction with 
an $e^+e^-$-collider. 
We have shown that in the partonic c.m.\ region that is relevant for the Higgs pair production process the obtained luminosity spectrum would be comparable to the one for an $e^-e^-$-based collider. 
For both the $e^-e^-$ and the $e^+e^-$ cases, the $\gamma\gamma$ collider mode 
could run in parallel to the $ee$-collider used for the Compton back-scattering at a second interaction region. 

The detailed assessment
of the accuracy with which the trilinear Higgs-boson self-coupling can be determined at a $\gamma\gamma$-collider of course requires 
dedicated experimental studies, taking into account all relevant backgrounds for the Higgs pair 
production process. Such studies are under way, see Ref.~\cite{slacpaper}  and Ref.~\cite{LinearColliderVision:2025hlt}. 
Since the interference patterns between the contributions involving the trilinear Higgs-boson self-coupling and the other diagrams contributing to the Higgs pair production processes
$\sigma(\gamma\gamma\to hh)$, $\sigma(gg\to hh)$, $\sigma(e^+e^-\to Zhh)$ and $\sigma(e^+e^-\to \nu\nu hh)$ 
are significantly different, it is obvious that the results from
a $\gamma\gamma$-collider will be highly complementarity to the measurements at the HL-LHC and at a high-energy $e^+e^-$ collider (at 550 GeV or 1 TeV).

The different $\gamma\gamma$-collider options discussed in 
this work all offer a very attractive programme for probing the Higgs potential
and driving innovation in accelerator and collider technologies. 
While further work on the technical feasibility and the detailed costing of  
$\gamma\gamma$-collider facilities is needed, 
significant cost savings can be expected for a $\gamma\gamma$-collider operating at 125~GeV for single Higgs production and at 280~GeV for Higgs pair production in comparison with an $e^+e^-$ collider at 250~GeV and 550~GeV, respectively, see e.g.\ the discussion in 
Ref.~\cite{Barklow:2023ess}, where the XCC at 125~GeV was compared with C$^3$-250. Thus, 
a $\gamma\gamma$-collider may prove to be the most economical way to probe the trilinear Higgs-boson self-coupling directly   
via the Higgs pair production process.

\vspace{1em}

\noindent\textit{\textbf{Acknowledgements:}}
 The authors would like to thank T.\ Barklow, S.\ Kanemura, T.\ Ohl, J.\ Reuter, A.\ Schwartzman and K.\ Yokoya
for helpful discussions.
We acknowledge support by the Deutsche Forschungsgemeinschaft (DFG, German Research Foundation) under Germany's Excellence Strategy --- EXC 2121 ``Quantum Universe'' --- 390833306. This work has been partially funded by the Deutsche Forschungsgemeinschaft (DFG, German Research Foundation) --- 491245950. 
M.B.\ is supported by the DFG Grant No.\ MO-2197/2-1. 
J.B.\ is supported by the DFG Emmy Noether Grant No.\ BR 6995/1-1.


\begin{thebibliography}{99}

\bibitem{Telnov:2020gwc}
V.~I.~Telnov,
JINST \textbf{15} (2020) no.10, P10028
[arXiv:2007.14003 [physics.acc-ph]].

\bibitem{Ginzburg}
I.F.~Ginzburg, G.L.~Kotkin, V.G.~Serbo, V.I.~Telnov,
``Production of High-Energy Colliding gamma gamma and gamma e Beams with a High Luminosity at Vlepp Accelerators''
JETP Lett. \textbf{34} (1981), 491-495;
I.F.~Ginzburg, G.L.~Kotkin, V.G.~Serbo, V.I.~Telnov,
``Colliding ge and gg beams based on the single-pass $e^\pm e^-$ colliders (VLEPP type)''
Nucl.Instrum.Meth. \textbf{205} (1983), 47-68; 
I.F.~Ginzburg, G.L.~Kotkin, S.L.~Panfil, V.G.~Serbo, V.I.~Telnov,
``Colliding gamma e and gamma gamma Beams Based on the Single Pass e+ e- Accelerators. 2. Polarization Effects. Monochromatization Improvement''
Nucl.Instrum.Meth.A. \textbf{219} (1984), 5-24

\bibitem{Barklow:2022vkl}
T.~Barklow, S.~Dong, C.~Emma, J.~Duris, Z.~Huang, A.~Naji, E.~Nanni, J.~Rosenzweig, A.~Sakdinawat and S.~Tantawi, \textit{et al.}
[arXiv:2203.08484 [hep-ex]].

\bibitem{Barklow:2023ess}
T.~Barklow, C.~Emma, Z.~Huang, A.~Naji, E.~Nanni, A.~Schwartzman, S.~Tantawi and G.~White,
JINST \textbf{18} (2023) no.07, P07028
[arXiv:2306.10057 [physics.acc-ph]].

\bibitem{Barklow:2024XCC}
Tim Barklow, 
“XCC status,” 
talk at LCWS 2024 (International Workshop on Future Linear Colliders), Tokyo, Japan, 8 July 2024.

\bibitem{ourpaper}
M.~Berger, J.~Braathen, G.~Moortgat-Pick, G.~Weiglein,
``Probing the Higgs potential via Higgs pair production at photon-photon colliders'', \textit{in preparation}.

\bibitem{slacpaper}
T.~Barklow, A.~Stratmann, et al.
``Higgs Self-Coupling Measurement with the XFEL Compton Collider (XCC)'',
\textit{in preparation}.

\bibitem{Hahn:2000kx}
T.~Hahn,
Comput. Phys. Commun. \textbf{140} (2001), 418-431
[arXiv:hep-ph/0012260 [hep-ph]].

\bibitem{Kublbeck:1990xc}
J.~Kublbeck, M.~Bohm and A.~Denner,
Comput. Phys. Commun. \textbf{60} (1990), 165-180

\bibitem{Hahn:1998yk}
T.~Hahn and M.~Perez-Victoria,
Comput. Phys. Commun. \textbf{118} (1999), 153-165
[arXiv:hep-ph/9807565 [hep-ph]].

\bibitem{Hahn:2016ebn}
T.~Hahn, S.~Pa{\ss}ehr and C.~Schappacher,
PoS \textbf{LL2016} (2016), 068
[arXiv:1604.04611 [hep-ph]].

\bibitem{Jikia:1992mt}
G.~V.~Jikia,
Nucl. Phys. B \textbf{412} (1994), 57-7

\bibitem{Asakawa:2008se}
E.~Asakawa, D.~Harada, S.~Kanemura, Y.~Okada and K.~Tsumura,
Phys. Lett. B \textbf{672} (2009), 354-360
[arXiv:0809.0094 [hep-ph]].

\bibitem{Asakawa:2010xj}
E.~Asakawa, D.~Harada, S.~Kanemura, Y.~Okada and K.~Tsumura,
Phys. Rev. D \textbf{82} (2010), 115002
[arXiv:1009.4670 [hep-ph]].

\bibitem{Bharucha:2020bhy}
A.~Bharucha, G.~Cacciapaglia, A.~Deandrea, N.~Gaur, D.~Harada, F.~Mahmoudi and K.~Sridhar,
JHEP \textbf{09} (2021), 069
[arXiv:2012.09470 [hep-ph]].

\bibitem{Moretti:2001zz}
M.~Moretti, T.~Ohl and J.~Reuter,
[arXiv:hep-ph/0102195 [hep-ph]].

\bibitem{Kilian:2007gr}
W.~Kilian, T.~Ohl and J.~Reuter,
Eur. Phys. J. C \textbf{71} (2011), 1742
[arXiv:0708.4233 [hep-ph]].

\bibitem{Kanemura:2004mg}
S.~Kanemura, Y.~Okada, E.~Senaha and C.~P.~Yuan,
Phys. Rev. D \textbf{70} (2004), 115002
[arXiv:hep-ph/0408364 [hep-ph]].

\bibitem{Kanemura:2017wtm}
S.~Kanemura, M.~Kikuchi, K.~Sakurai and K.~Yagyu,
Phys. Rev. D \textbf{96} (2017) no.3, 035014
[arXiv:1705.05399 [hep-ph]].

\bibitem{Bahl:2022jnx}
H.~Bahl, J.~Braathen and G.~Weiglein,
Phys. Rev. Lett. \textbf{129} (2022) no.23, 23
[arXiv:2202.03453 [hep-ph]].

\bibitem{Biekotter:2022kgf}
T.~Biek{\"o}tter, S.~Heinemeyer, J.~M.~No, M.~O.~Olea-Romacho and G.~Weiglein,
JCAP \textbf{03} (2023), 031
[arXiv:2208.14466 [hep-ph]].

\bibitem{Bahl:2023eau}
H.~Bahl, J.~Braathen, M.~Gabelmann and G.~Weiglein,
Eur. Phys. J. C \textbf{83} (2023) no.12, 1156
[erratum: Eur. Phys. J. C \textbf{84} (2024) no.5, 498]
[arXiv:2305.03015 [hep-ph]].

\bibitem{Braathen:2025qxf}
J.~Braathen, S.~Heinemeyer, A.~P.~Arnay and A.~Verduras Schaeidt,
[arXiv:2507.02569 [hep-ph]].

\bibitem{ATLAS:2024ish}
G.~Aad \textit{et al.} [ATLAS],
Phys. Rev. Lett. \textbf{133} (2024) no.10, 101801
[arXiv:2406.09971 [hep-ex]].


\bibitem{LinearColliderVision:2025hlt}
D.~Atti{\'e} \textit{et al.} [Linear Collider Vision],
[arXiv:2503.19983 [hep-ex]].

\bibitem{Berggren:2025fpw}
M.~Berggren, B.~Bliewert, J.~List, D.~Ntounis, T.~Suehara, J.~Tian, J.~M.~Torndal and C.~Vernieri,
[arXiv:2509.14148 [hep-ex]].

\bibitem{LHCHiggsCrossSectionWorkingGroup:2016ypw}
D.~de Florian \textit{et al.} [LHC Higgs Cross Section Working Group],
CERN Yellow Rep. Monogr. \textbf{2} (2017), 1-869
[arXiv:1610.07922 [hep-ph]].

\bibitem{Chen:1994jt}
P.~Chen, G.~Horton-Smith, T.~Ohgaki, A.~W.~Weidemann and K.~Yokoya,
Nucl. Instrum. Meth. A \textbf{355} (1995), 107-110

\bibitem{Bechtel:2006mr}
F.~Bechtel, G.~Klamke, G.~Klemz, K.~Monig, H.~Nieto, H.~Kluge, A.~Rosca, J.~Sekaric and A.~Stahl,
Nucl. Instrum. Meth. A \textbf{564} (2006), 243-261
[arXiv:physics/0601204 [physics]].


\bibitem{Kawada:2012uy}
S.~i.~Kawada, N.~Maeda, T.~Takahashi, K.~Ikematsu, K.~Fujii, Y.~Kurihara, K.~Tsumura, D.~Harada and S.~Kanemura,
Phys. Rev. D \textbf{85} (2012), 113009
[arXiv:1205.5292 [hep-ph]].




\end{thebibliography}
\end{document}